\documentclass[aps,prx,twocolumn,amsmath,amssymb,superscriptaddress,floatfix]{revtex4-2}
\usepackage{mathtools}
\usepackage{multirow}
\usepackage{xcolor}
\usepackage{bm}
\usepackage{amsfonts,amssymb,amsmath}
\usepackage{graphicx,dcolumn,bm,xcolor,braket,slashed}
\usepackage{times} 
\usepackage{comment}
\usepackage{array}
\usepackage{textcomp}
\usepackage[normalem]{ulem}
\usepackage{dsfont}
\usepackage{hyperref}
\usepackage{float}

\usepackage[title,toc,titletoc,page]{appendix}

\newcommand{\be}{\begin{eqnarray}}
\newcommand{\ee}{\end{eqnarray}}
\newcommand{\bbm}{\begin{bmatrix}}
\newcommand{\ebm}{\end{bmatrix}}
\newcommand{\bpm}{\begin{pmatrix}}
\newcommand{\epm}{\end{pmatrix}}

\begin{document}

\title{Geometric Approach to Zero-Memory Quantum Dot Reservoir Computing}

\author{Bongsu Kim}
\email{kbswhy@cau.ac.kr}
\affiliation{Department of Physics, Chung-Ang University, 06974 Seoul, Republic of Korea}

\author{Oscar Lee}
\affiliation{London Centre for Nanotechnology, University College London, London WC1H 0AH, United Kingdom}

\author{Sangjun Jeon}
\affiliation{Department of Physics, Chung-Ang University, 06974 Seoul, Republic of Korea}

\author{Kun Woo Kim}
\email{kunx@cau.ac.kr}
\affiliation{Department of Physics, Chung-Ang University, 06974 Seoul, Republic of Korea}

\begin{abstract}
Physical reservoir computing offers an energy-efficient alternative to conventional neural networks, where the material-specific intrinsic memory capacity in a physical system plays an indispensable role. Substituting temporal memory with spatial degrees of freedom, we demonstrate that the memory capacity can be created extrinsically in systems with no intrinsic memory by exploiting the computational space-time tradeoff. 
Our approach utilizes multidimensional input nodes to function as a spatial memory axis, thereby replacing the dependency on intrinsic history-dependent dynamics in the reservoir. 
Our scheme is validated in a multi-terminal quantum dot system, whose discrete energy levels provide strong nonlinearity and complexity crucial for reservoir computing, while its short relaxation time leaves no room for intrinsic memory.
Numerically, the quantum dot reservoir with a tunable extrinsic memory shows high performance on both chaotic future prediction and nonlinear transformation tasks. 
Furthermore, from the analysis of quantum state trajectory acquired from task operations, the geometric understanding of the extrinsic memory capacity, nonlinearity, and complexity is provided and their correlations are systematically investigated. 

\end{abstract}

\maketitle
\newpage
\section{Introduction} 

The rapid expansion of artificial intelligence, particularly large-scale deep learning models, has increased computational demand and energy consumption in data centers, raising concerns about the sustainability and cost of training and deploying such models. In response to this challenge, Physical Reservoir Computing (PRC) has emerged as a compelling alternative framework \cite{jaeger2001echo, dambre2012information}. Unlike standard neural networks that require energy-intensive optimization processes across all layers, PRC exploits the rich dynamics of a physical substrate—such as optical \cite{nakajima2021scalable, van2017advances}, spintronic \cite{taniguchi2022spintronic, zhou2025harnessing, chen2025spintronic}, or memristor systems \cite{du2017reservoir, chen2025inse, zhong2021dynamic, sun2021sensor}—to project inputs into a high-dimensional state space, requiring training only for the final output layer \cite{jaeger2002tutorial, butcher2013reservoir, torrejon2017neuromorphic, barbosa2021symmetry, milano2022materia}. More recently, this has been extended into quantum regime, where quantum reservoir computing utilizes the exponentially large Hilbert space \cite{Mujal2021, Suzuki2022, PhysRevLett.127.100502, Mujal2023, PRXQuantum.5.040325}. However, the efficacy of a physical reservoir in processing time-series data, such as chaotic forecasting or speech recognition, depends critically on its capacity for memory \cite{jaeger2001short, lee2024task, zhong2021dynamic, du2017reservoir}. Conventionally, this memory is assumed to be an intrinsic property of a physical system, often arising from relaxation processes or hysteresis that allow the system's current state to retain echoes of past inputs. Consequently, the search for suitable reservoir materials has largely been confined to physical systems that naturally exhibit this memory property, limiting the scope of viable candidates for reservoir computing \cite{tanaka2019recent, nakajima2020physical, liang2024physical}. 

To overcome this material-given property, several approaches have been proposed to manipulate the memory effect extrinsically. One class enhances the memory supplied by the reservoir itself, such as using time-delayed feedback \cite{appeltant2011information, goldmann2020deep, tavakoli2024boosting, larger2017high, haynes2015reservoir, argyris2021fast}, optical cavities \cite{brunner2013parallel, vandoorne2014experimental, nakajima2021scalable, van2017advances}, and rich phase diagrams of magnetic systems \cite{lee2024task, lee2023, gartside2022reconfigurable, torrejon2017neuromorphic, kanao2019reservoir}. Although effective at many tasks, their performance hinges on the system operating near the edge of chaos and its internal relaxation timescale matching the input signal rate. These constraints arise because memory is still realized through the history-dependent intrinsic dynamics of the reservoir. 
A second class acts outside the reservoir, in the input layer \cite{Jaurigue2024, Fleddermann2025} or the readout layer \cite{DelFrate2021, Carroll2022, Picco2023, PhysRevResearch.1.033030}, using delayed copies of the signal or of the response. These approaches bypass the reliance on inherent reservoir dynamics and allow the required memory to be engineered externally. However, most of these works still employ reservoirs with memory and treat the extrinsic scheme as a supplement to it. The case of reservoirs without memory has also been tested \cite{Jaurigue2024, Fleddermann2025}, but only as a limiting control, which leaves a natural question open: Can a memoryless system serve as an efficient reservoir computer if its memory capacity is supplied entirely from outside?

In this work, we answer this question by showing that a memoryless quantum dot can serve as an efficient reservoir even on a forecasting task. Quantum dots are promising nanodevices, and experimental advancements for their control have reached a mature stage \cite{clifford2009fast, yamahata2019picosecond, kagan2016building, picosecond2017czerniuk, real2021nakajima}. In particular, quantized and irregular energy levels endow them with rich nonlinearity and complexity. However, despite these advantages, its application to the PRC is restricted by a lack of intrinsic memory due to their fast relaxation time. We overcome this critical limitation by supplying the memory from outside the system, based on the computational principle of the space-time tradeoff: we inject the input history in parallel across multiple sites, so that spatial degrees of freedom take over the role normally played by internal dynamics \cite{savage1998models, savage1978space, hellman1980cryptanalytic}. Our results show that the quantum dot can be tuned as a versatile reservoir, achieving high performance on both nonlinear transformation and the chaotic Mackey-Glass time series prediction task \cite{mackey1977oscillation}, a benchmark known to require high memory capacity \cite{love2023spatial, lee2024task, gartside2022reconfigurable, tavakoli2024boosting}. The entire computation is fully conducted in discrete steps without relying on actual time, since the input rate is no longer tied to the hardware's relaxation timescale \cite{Gauthier2021, 1380068}, which offers the possibility of performing \textit{parallel} physical reservoir computations. 

\begin{figure*}
    \centering
    \includegraphics[width=1.0\linewidth]{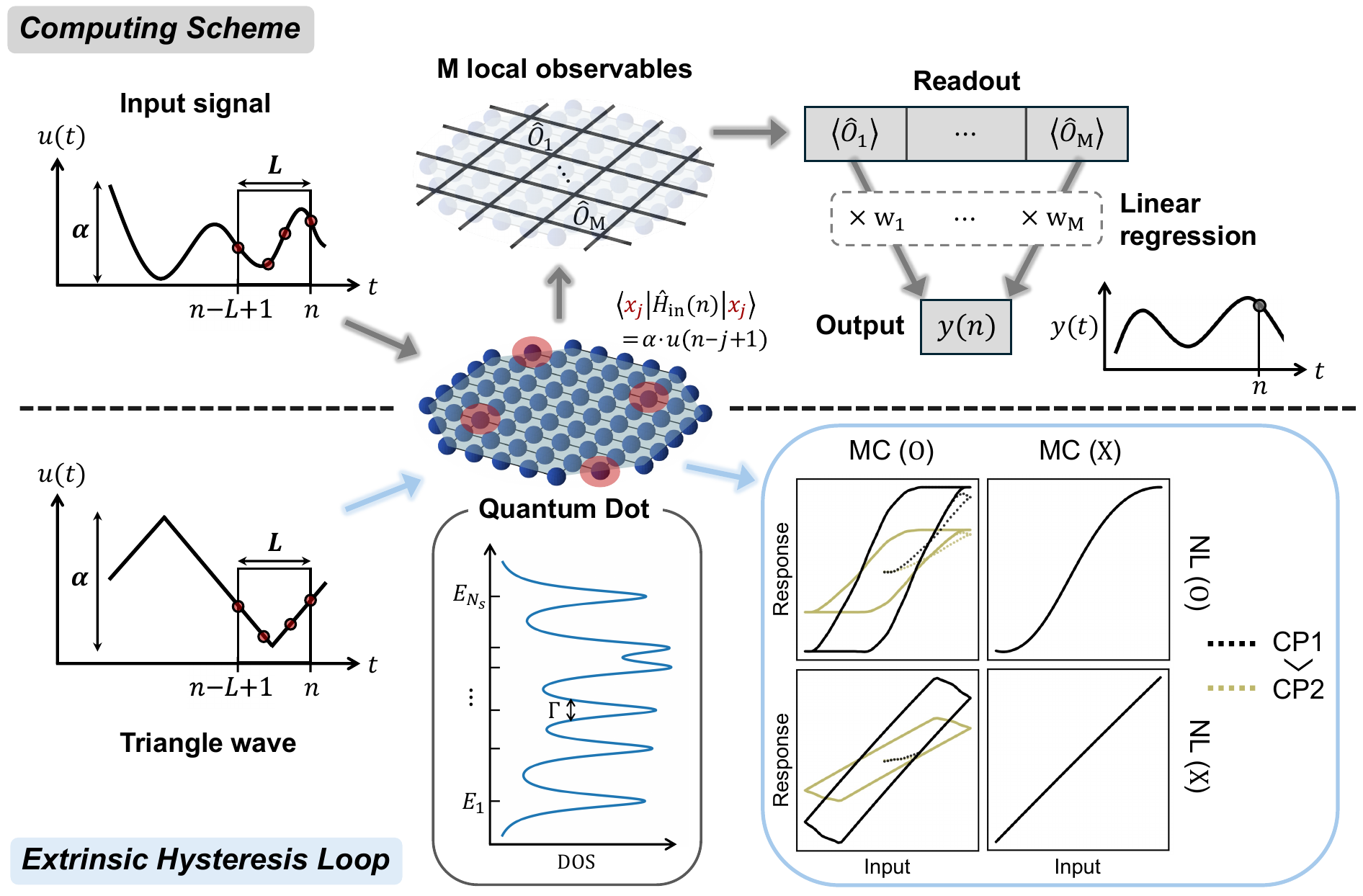}
    \caption{\textbf{Quantum dot reservoir computing scheme with extrinsic hysteresis loop.} Top: We use a single quantum dot as a physical reservoir. A discrete input time series, derived from the continuous input signal $u(t)$, is applied to $L$ input nodes (here we set $L=4$), which correspond to distinct sites on the system (denoted as $x_j$ and marked in red). Specifically, at every time step $n$, the input consists of the signal history within an $L$-step window $[u(n), \dots, u(n-L+1)]$, scaled by $\alpha$, injected in parallel across the $L$ spatial nodes as onsite potentials to form the input Hamiltonian $\hat{H}_{\text{in}}(n)$. We then measure local observables $\hat{O}_i \ (i=1,\dots, M)$ across different regions to obtain $M$ readout values. Using these, we perform linear regression to train the weights for each readout node, finally obtaining a single output value $y(n)$. 
    Bottom: As a consequence of both the extrinsic memory induced by the input injection scheme and the inherent nonlinear properties of the quantum dot (reflected in its density of states with energy broadening $\Gamma$), a hysteresis loop emerges in the input-response space. To visualize this loop structure, a triangle wave is used as the input. By tuning the control parameters $L$ and $\alpha$, we confirm that the reservoir's properties---characterized by memory capacity (MC), nonlinearity (NL), and complexity (CP)---can be tuned, and therefore the shape of the hysteresis loop changes systematically, with each reservoir metric leaving a distinct geometric signature: MC determines the hole, NL introduces curvature, and CP determines the response range.}
    \label{fig.scheme}
\end{figure*}

Analyzing the trajectory of the quantum dot state evolving in computation steps, we provide a geometric understanding of extrinsically created memory capacity and its correlation with nonlinearity and complexity for physical reservoir computation. That is, within the visualized  Hilbert space, a hysteresis-like loop structure is developed as the extrinsic memory grows, and its size and uneven shape directly manifest the magnitude of the three performance measures. The topological robustness of the PRC performance can be demonstrated using 1-dimensional input signals, however, with a different number of incommensurate frequencies which explores a higher dimensional synthetic space. 

The rest of the paper is structured as follows. 
In Sec.~\ref{sec.extrinsic}, we introduce the reservoir computing procedure using a multi-terminal quantum dot model and the input injection scheme that realizes memory spatially.
In Sec.~\ref{sec.tunable}, we evaluate the three reservoir metrics and show that they are tunable through parameters in quantum dot devices.
In Sec.~\ref{sec.task}, we benchmark the memoryless reservoir on a nonlinear transformation and a future prediction task, and examine how the control parameters govern it.
In Sec.~\ref{sec.homology}, we analyze the geometry that the reservoir unfolds in the visualized Hilbert space via persistent homology, relating the size and distortion of the emergent loop to the reservoir metrics. 
Finally, we discuss the advantage of physical nonlinearity, the use of geometry as a reservoir diagnostic, and experimental realization in Sec.~\ref{sec.discuss}.


\section{Extrinsically introduced Memory Capacity}
\label{sec.extrinsic}

Memory refers to the capacity to store and retrieve information about past events. It implies that its current state and its responses are not solely determined by present inputs, but also by its past history. This history dependence is the fundamental characteristic of a system possessing the memory capacity. Physical systems can contain intrinsic memories~\cite{kittel1946theory, tokura2014multiferroics} and they are reflected in the static and dynamical properties. 
For instance, when such a ferromagnetic material is driven by an external magnetic field, $H(t)\sim \sin(\omega t)$, its history-dependency causes the magnetic response to exhibit a temporal lag, yielding $M(t)\sim\sin(\omega t - \phi)$. In addition, the inherent nonlinearity of the magnetization process introduces higher-order harmonic distortions into this response. The combination of the temporal delay and this nonlinear response gives rise to the well-known magnetic hysteresis loop in the input-response space \cite{jiles1984theory} (see Fig.~\ref{fig.scheme}).
Taking advantage of physical systems with intrinsic memory, it is possible to perform complex information processing tasks directly within the physical hardware such as prediction of pseudo random function and forecasting tasks \cite{tanaka2019recent, nakajima2020physical, liang2024physical}. 

While intrinsic memory in a physical system is considered a prerequisite for conventional PRC, our work establishes an alternative approach. We demonstrate that this reliance can be circumvented by engineering an input injection scheme that explicitly encodes the signal's history into parallel, time-delayed signals, thereby forming an artificial hysteresis loop, as illustrated in Fig.~\ref{fig.scheme}. We model a single quantum dot using a finite-sized tight-binding Hamiltonian: 
\begin{equation}
    \hat{H}_{\text{sys}} = \sum_{i, j=1}^{N_s} J_{ij}(c_i^\dagger c_j + \text{h.c.}) + \sum_{i=1}^{N_s} \epsilon_i c_i^\dagger c_i.
\end{equation}
Here, $c_i (c_i^\dagger)$ is spinless annihilation (creation) operator at site $i$, $J_{ij}$ is the internal hopping strength, $\epsilon_i$ is the onsite energy, and $N_s$ is the total number of sites. While a quantum dot inherently exhibits memory effects within its relaxation time scale, our computing scheme operates within a discrete-time framework, assuming that the system's internal dynamics are fast enough that the quantum dot fully relaxes to equilibrium within each discrete time step $n$. Therefore, we treat it as a memoryless system. We use a random matrix for $\hat{H}_{\text{sys}}$ to demonstrate the general applicability of our scheme.
At each time step $n$, to induce the memory extrinsically, multi-delayed input signals $u(n), \dots, u(n-L+1)$ are injected into $L$ input nodes as an additional onsite potential. The corresponding Hamiltonian $\hat{H}(n)$ is expressed as $\hat{H}(n) = \hat{H}_{\text{sys}} + \hat{H}_{\text{in}}(n)$, where the input term is defined as:
\begin{equation}
    \hat{H}_{\text{in}}(n) = \alpha \left[ u(n) c_{x_1}^\dagger c_{x_1} + \dots + u(n-L+1) c_{x_L}^\dagger c_{x_L} \right],
\label{eq:permute}
\end{equation}
where $\alpha$ is the input scaling factor and $x_j \ (j=1, \dots, L)$ is the lattice index of the $j$-th input node. This expansion explicitly maps the zero-lag signal $u(n)$ to site $x_1$, while mapping the signal with a lag of $L-1$ to site $x_L$ in parallel, thereby utilizing the spatial nodes as a temporal memory axis. 
Besides this specific input mapping, the multi-delayed signals can be mapped in other ways. We introduce and compare four different schemes in Appendix~\ref{appendix.mappings}, demonstrating that the mapping of Eq.~\ref{eq:permute} is optimal for generating memory.

The internal state of the physical reservoir is represented by the wavefunction $|\psi_k(n)\rangle \in \mathbb{C}^{N_s}$, where $k$ denotes the eigenstate index. This describes an isolated system without any environmental interaction. When the system is coupled to an environment, such as measurement apparatus, it transitions into a statistically mixed state described by a density matrix $\rho(n)=\sum_k p_k \ket{\psi_k(n)} \bra{\psi_k(n)}$, where $p_k$ is the classical probability of the system being in the state $\ket{\psi_k(n)}$. Our model accounts for this environmental interaction, thereby providing a more accurate reflection of experimental conditions.

The information processed by the reservoir is extracted through a set of $M$ readout nodes, defined by partitioning the lattice into $M$ mutually disjoint regions $\{\mathcal{R}_m\}_{m=1}^M$, where $\mathcal{R}_m \subset \{1,\dots,N_s\}$ denotes the set of sites belonging to the $m$-th region. In our experiments, we choose $M=N_s$. Given an arbitrary Hermitian measurement operator $\hat{O}$ acting on the system's Hilbert space, we define the local observable for the $m$-th region as $\hat{O}_m = \hat{P}_m \hat{O} \hat{P}_m$, where $\hat{P}_m = \sum_{i \in \mathcal{R}_m} |i\rangle\langle i|$ is the projection operator onto the subspace spanned by the sites in $\mathcal{R}_m$. The $m$-th component of the readout vector $\mathbf{r}(n; E)$, which corresponds to the expectation value of the local observable $\braket{\hat{O}_m}$ at a fixed energy $E$ at time step $n$, is then given by:
\begin{equation}
    r_m(n; E) = \frac{1}{\pi} \Im \left[ \sum_k \frac{\braket{\psi_k(n)|\hat{O}_m|\psi_k(n)}}{E - E_k(n) - i\Gamma} \right],
\label{eq:readout}
\end{equation}
where $E_k(n)$ is the eigenenergy corresponding to the state $\ket{\psi_k(n)}$, and $\Gamma$ is an energy broadening factor that represents environmental coupling. In the limit $\Gamma \rightarrow 0^+$, this equation recovers the delta function representation, as $\Im[(E-E_k(n)-i\Gamma)^{-1}] \rightarrow \pi \delta(E-E_k(n))$.
Collecting this readout vector over the full input duration $N$, we construct the readout matrix $R \in \mathbb{R}^{N \times M}$, where the $n$-th row corresponds to the system response at time $n$. For notational simplicity, we drop the energy dependence $E$ hereafter:
\begin{equation}
    R = 
    \begin{bmatrix}
        r_1(1) & r_2(1) & \cdots & r_M(1) \\
        r_1(2) & r_2(2) & \cdots & r_M(2) \\
        \vdots & \vdots & \ddots & \vdots \\
        r_1(N) & r_2(N) & \cdots & r_M(N)
    \end{bmatrix}.
    \label{eq:reservoir_matrix}
\end{equation}
The final output signal $y(n)$ is obtained via a linear combination of the readout components, mediated by an output weight vector $\mathbf{w} \in \mathbb{R}^{M}$, given by $ y(n) = \mathbf{r}(n) \cdot \mathbf{w} = \sum_{m=1}^M r_m(n) \text{w}_m$. In searching for the optimal weights, we employ Lasso regression (L1 regularization) \cite{tibshirani1996regression} against a desired target signal. The training is performed on the first 70\% of the data (spanning $N_{\text{train}}=0.7N$ time steps), while the remaining 30\% ($N_{\text{test}}$) serves as the testing set.

Figure~\ref{fig.scheme} (bottom) illustrates that the proposed scheme, along with the quantum dot's nonlinearity, can engineer the hysteresis loop extrinsically. We use a triangle wave as the input to visualize the loop structure and plot the reservoir response against the input. Note that this response can be either the system state or the readout vector. By tuning the reservoir metrics through control parameters $L$ and $\alpha$ (see Sec.~\ref{sec.tunable}), the shape of the hysteresis loop changes: MC governs the opening of the hole, NL introduces curvature, and CP determines the response range. These geometric features provide an intuitive picture of the reservoir's computational richness, motivating a deeper analysis of the quantum state trajectories in Sec.~\ref{sec.homology}.

From a computer science perspective, our approach is a physical implementation of the space-time tradeoff, a principle where memory (space) and processing steps (time) are treated as interchangeable computational resources \cite{carroll2019network, wolpert2019space, fujii2017harnessing, corlett2025speeding}. 
In conventional reservoir computing using a single input node, retaining input information requires the system to possess fading memory, so that the observable state $\mathbf{r}(n)$ depends on its previous state $\mathbf{r}(n-1)$ through internal dynamics within the relaxation timescale. This temporal retention corresponds to the computational time cost.
In our proposed scheme, we utilize multidimensional input nodes to establish a spatial memory axis, spending more space so that the tradeoff reduces the retention time required within the reservoir. For a memoryless reservoir with $L$ distinct input nodes, the temporal history of the signal is embedded into the input Hamiltonian $\hat{H}_{\text{in}}(n)$ and injected across the spatial nodes. Because the system is intrinsically memoryless, the resulting readout $\mathbf{r}(n)$ represents the steady-state response to the inputs currently being applied, and the past state is completely overwritten when the next set arrives. Nevertheless, since the current injection already carries the history, past inputs up to $L$ time steps remain accessible to the computation. Setting $L$ to match the desired memory timescale therefore substitutes intrinsic temporal memory with extrinsic spatial memory, significantly expanding the range of physical systems applicable to PRC.

\section{Tunable Reservoir Metrics}
\label{sec.tunable}

Reservoir metrics provide a numerical representation of the reservoir's properties for a given input signal \cite{love2023spatial}. We use an independent and identically distributed (i.i.d.) random input signal $u(n)$ drawn from a uniform distribution in the range $[0, 1]$, ensuring that the measured metrics arise solely from the system. First, memory capacity (MC) quantifies the ability to retain information about past inputs. Here, we focus on the linear memory capacity, which measures how well past information can be retrieved using a linear regression. Specifically, we evaluate how accurately a previous input signal $u(n-\tau)$, delayed by $\tau$ steps, can be reconstructed from the current observable state $\mathbf{r}(n)$. The linear estimator is given by $\hat{u}(n-\tau) = \mathbf{r}(n) \cdot \mathbf{w}_{\tau}$, where a distinct weight vector $\mathbf{w}_\tau \in \mathbb{R}^{M}$ is trained for each $\tau$. The total MC is then defined as the sum of the squared correlation coefficients between the true past inputs $\mathbf{u}_{\tau}$ and the estimated values $\hat{\mathbf{u}}_{\tau}$, both of length $N_{\text{test}}$.
\begin{equation}
    \text{MC} = \sum_{\tau=1}^{k} \frac{\text{cov}^2(\hat{\mathbf{u}}_{\tau},\, \mathbf{u}_{\tau})}{\sigma^2(\hat{\mathbf{u}}_{\tau})\sigma^2(\mathbf{u}_{\tau})},
\label{eq:MC}
\end{equation}
where $\text{cov}(\cdot, \cdot)$ denotes the covariance and $\sigma^2(\cdot)$ represents the variance. The upper limit $k$ must be chosen to be larger than the relaxation time of the system to get a convergent value, meaning $k > L$ in our case. Because the inequality $|\text{cov}(X, Y)| \leq \sqrt{\sigma^2(X) \sigma^2(Y)}$ holds for arbitrary variables $X$ and $Y$ with finite variance, each term in the summation ranges from zero (no correlation) to one (perfect prediction).

Second, nonlinearity (NL) measures the degree to which the reservoir's dynamics deviate from a purely linear response. This is quantified by inverting the regression process used for MC, where the true readout signal $r_m(n)$ is approximated using a linear combination of recent inputs. The estimator is given by $\hat{r}_m(n) = \mathbf{u}_k(n) \cdot \mathbf{w}_{m}$, where $\mathbf{u}_k(n) = (u(n), \dots, u(n-k+1))$ is the input history of length $k$, and the weight $\mathbf{w}_m \in \mathbb{R}^k$ is trained for each readout node. The overall NL is then calculated by averaging over $M$ nodes.
\begin{equation}
        \text{NL} = 1-\frac{1}{M}\sum_{m=1}^{M} \frac{\text{cov}^2(\hat{\mathbf{r}}_m,\, \mathbf{r}_m)}{\sigma^2(\hat{\mathbf{r}}_m)\sigma^2(\mathbf{r}_m)}.
\end{equation}
A value of zero (one) indicates that the relationship between input and readout is purely linear (nonlinear).

Lastly, complexity (CP) determines the ability of the reservoir to generate a diverse set of readout signals. While the NL and MC calculations evaluate the relationship between input and readout, CP is determined using only the readout matrix $R$. We compute this metric by applying singular value decomposition to obtain singular values $\sigma_i$ in descending order, where $i=1,\, \dots, Q$ and $Q = \min\{N, M\}$. By normalizing these values, we construct a probability distribution $p_i$, and the Shannon entropy $H$ of this distribution is calculated as $H(p_1,\, \dots,\, p_Q) = -\sum_{i=1}^Q p_i \log p_i$. Then CP is defined as the effective rank of the matrix $R$.
\begin{equation}
    \text{CP} = \text{erank}(R) = \exp \{H(p_1,\, ...,\, p_Q)\}.
\end{equation}
CP takes a real value in the interval $[1, Q]$, where a higher complexity indicates that the reservoir spans a higher-dimensional latent space. Throughout this work, all reservoir metrics are calculated using i.i.d input of length $N=1,000$.
\begin{figure*}
    \centering
    \includegraphics[width=1.0\linewidth]{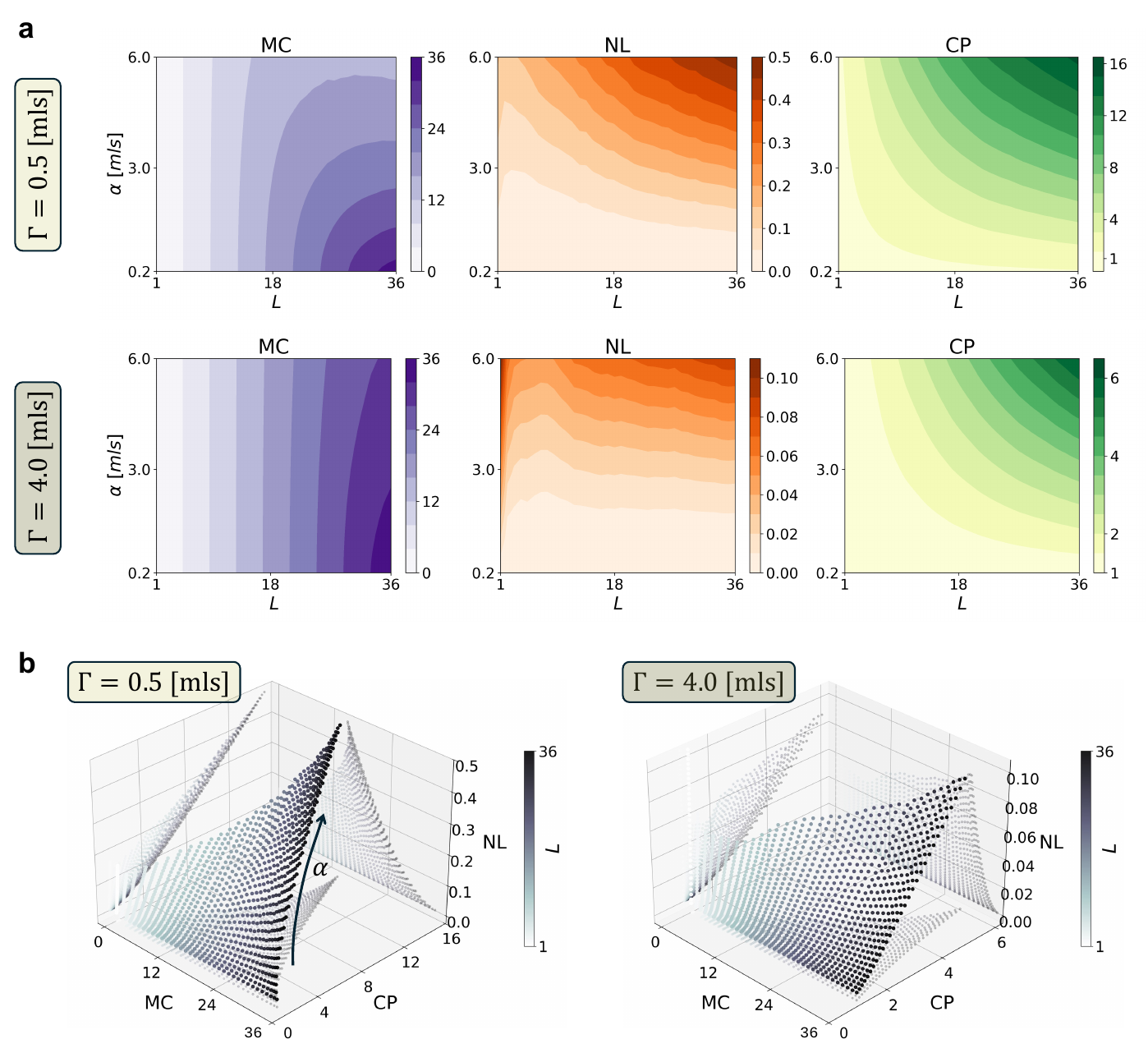}
    \caption{\textbf{Tunable reservoir metrics and their correlations across the control parameter space.} (a) Heatmaps of memory capacity (MC), nonlinearity (NL), and complexity (CP) as functions of the number of input nodes $L$ and the input scaling factor $\alpha$, computed using the identity measurement operator $\hat{O}=\hat{I}$ and averaged over $20$ random seeds of the system Hamiltonian $\hat{H}_{\text{sys}}$. The panels compare low $(\Gamma=0.5)$ and high $(\Gamma=4.0)$ energy broadening regimes, where contour lines indicate the distinct scaling behaviors driven by the two control axes. Both $\Gamma$ and $\alpha$ are expressed in units of the mean level spacing (mls) of $\hat{H}_{\text{sys}}$.
(b) Three-dimensional representations of the reservoir metric space using the same data from (a), with individual points colored according to $L$. These plots illustrate the correlation shifts induced by $\Gamma$ and demonstrate that the parameter space $(L, \alpha)$ provides navigation paths that bypass the conventional trade-off between MC and NL.}
    \label{fig.tunability}
\end{figure*}

Tunability of the three reservoir metrics is crucial to perform various computational tasks for a fixed reservoir. In Fig.~\ref{fig.tunability}a, heatmaps of MC, NL, and CP are shown over the two-dimensional control parameter space ($L, \alpha$) for $\Gamma=0.5$ and $\Gamma=4.0$, using the quantum dot with $N_s=36$. Throughout this work, the parameters $\Gamma$ and $\alpha$ are expressed in units of the mean level spacing (mls) of the system Hamiltonian. All three metrics exhibit clear and systematic tunability across the parameter space, yet they are governed by distinct control axes. For $\Gamma=0.5$, MC increases strongly with $L$ but decrease monotonically with $\alpha$, whereas NL is predominantly driven by $\alpha$ with a weaker dependence on $L$. CP responds positively to both parameters jointly, reaching its maximum when both $L$ and $\alpha$ are large.
These behaviors can be understood in terms of how $L$ and $\alpha$ perturb the system. Increasing $L$ assigns a larger number of input node sites to receive a longer input history, while increasing $\alpha$ means stronger input injection. From the readout expression in Eq.~\ref{eq:readout}, either a larger $\alpha$ or a larger $L$ induces stronger perturbations to the system, causing the eigenenergies $E_k(n)$ to shift more widely relative to the fixed readout energy $E$. Consequently, a greater number of eigenmodes contribute to the readout. This enriches the nonlinear character of $\mathbf{r}(n)$, as the nonlinearity originates from the discretized energy levels of the quantum dot, and diversifies the readout set, but it simultaneously makes it harder to linearly isolate and reconstruct past inputs $u(n-\tau)$, which underlies the conventional trade-off between MC and NL \cite{dambre2012information, inubushi2017reservoir}.

For $\Gamma=4.0$, the increased broadening changes the scaling behavior of the reservoir metrics. A larger $\Gamma$ causes neighboring eigenstates to overlap more significantly, so the readout receives mixed contributions from a wider range of eigenmodes, which reduces the number of linearly independent readout signals. This effect is visible in the contour structures of MC and NL. Compared to $\Gamma=0.5$, the contour lines at $\Gamma=4.0$ shift toward the low-$\alpha$ regime, indicating that the state overlap by $\Gamma$ diminishes the state mixing effect by $\alpha$. Consequently, NL becomes confined to a smaller region, while MC becomes more uniform across $\alpha$, losing much of its sensitivity to the input scaling. CP, however, behaves differently due to its mathematical definition. Although the maximum value of CP is reduced due to the loss of effective dimensionality, the shape of its contour lines remains relatively unchanged compared to the low-$\Gamma$ case. This is because CP is determined by the relative distribution of singular values of the readout matrix $R$. As $\Gamma$ increases the overlap, it rescales the entries of $R$ by a similar factor. Because CP is calculated using the normalized singular values $p_i$, this rescaling by $\Gamma$ cancels out and therefore the relative distribution of singular values is preserved, leaving the contour structure of CP intact.

Highlighting the correlation between the reservoir metrics, the data from Fig.~\ref{fig.tunability}a is replotted as a three-dimensional reservoir metric space in Fig.~\ref{fig.tunability}b, with each point colored according to $L$. For $\Gamma=0.5$, NL and CP are highly positively correlated with each other, as both are strongly driven by increases in $\alpha$. In contrast, MC and NL typically exhibit an inverse relationship along the $\alpha$ axis, where higher $\alpha$ boosts NL but reduces MC.
As the energy broadening increases to $\Gamma=4.0$, the state overlap shifts these correlations. NL almost loses its $L$ dependency, MC loses its $\alpha$ dependency, but CP maintains the dependency on both. Consequently, MC and NL become uncorrelated in this high-$\Gamma$ regime, at the cost of a reduced range of NL.
Taken together, these results reveal that we can tune the reservoir metrics more flexibly by using extrinsic memory, which introduces $L$ as an additional control axis. Importantly, this makes it possible to violate the conventional trade-off between MC and NL, which typically constrains reservoirs to sacrifice one for the other. Even in the low-$\Gamma$ regime, we can navigate the parameter space $(L, \alpha)$ to decouple temporal memory retention from nonlinear dynamics.

\section{Task Performance}
\label{sec.task}

Here, we test the inherent nonlinearity of quantum dot and the efficacy of the extrinsic memory via two different tasks: a nonlinear transformation and a chaotic time-series future prediction, both of which require sufficient NL and MC. In nonlinear transformation task, we inject a sine wave as an input and apply linear regression to the reservoir readouts to target a sawtooth wave, which has a nonlinear relationship with the sine input. For the prediction task, we employ the Mackey-Glass (MG) time series, a widely explored benchmark in reservoir computing field, obtained by solving the following equation:
\begin{equation}
    \frac{dx}{dt} = -ax(t) + \frac{bx(t-d)}{1+x^{c}(t-d)},
\end{equation}
where we set the standard parameters to $a=0.1, b=0.2, c=10, d=17$. We inject this chaotic signal and forecast its 20-steps-ahead signal. The task performance is evaluated using the mean squared error (MSE) between the true target signal $\mathbf{y}$ and the predicted output $\hat{\mathbf{y}}$. Specifically, over the test set, this is computed as $\text{MSE}(\mathbf{y}, \hat{\mathbf{y}}) = \frac{1}{N_{\text{test}}}\sum_{i=1}^{N_{\text{test}}}(y_i-\hat{y}_i)^2$.

\begin{figure*}
    \centering
    \includegraphics[width=1.0\linewidth]{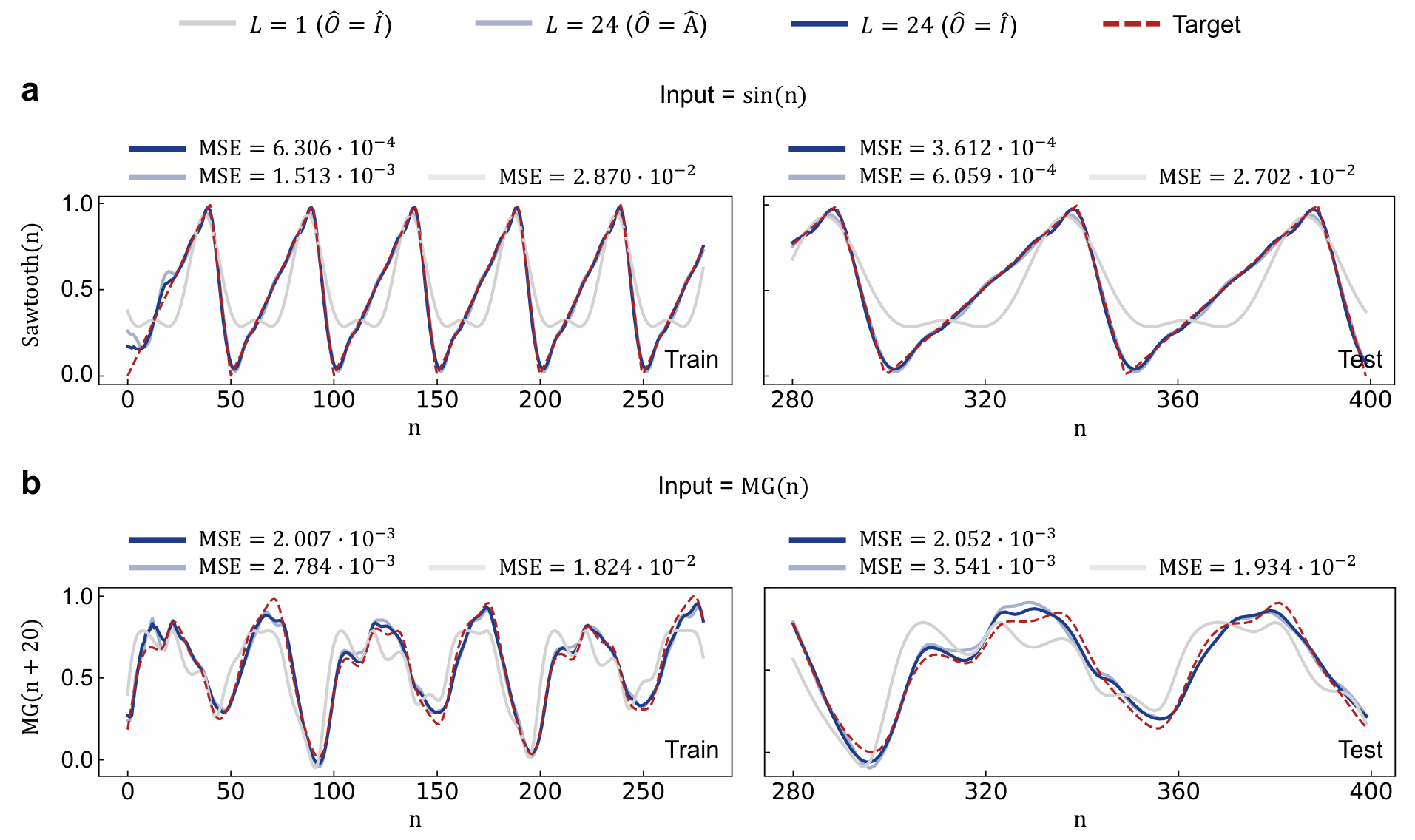}
    \caption{\textbf{Task performance of the quantum dot reservoir.} (a) Nonlinear transformation task, where a sine input is mapped to a sawtooth target. (b) Mackey-Glass 20-step-ahead prediction task. In both panels, the left and right subplots show the train $(N_{\text{train}}=280)$ and test $(N_{\text{test}}=120)$ sets respectively, and three configurations are compared: $L=1$ with $\hat{O}=\hat{I}$ (light gray), $L=24$ with $\hat{O}=\hat{I}$ (dark blue), and $L=24$ with $\hat{O}=\hat{A}$ (light blue), where $\hat{A}$ is a randomly generated observable, against the target signal (red dashed). The reservoir parameters are $N_s=36, \alpha=3.0,$ and $\Gamma=0.5$. We evaluate each task performance via mean squared error (MSE), where a smaller value indicates better performance.}
    \label{fig.task}
\end{figure*}

The task performance of the reservoir with $N_s=36$ sites and input length $N=400$ is plotted in Fig.~\ref{fig.task}, with $\alpha=3.0$ and $\Gamma=0.5$ fixed. The first 70\% of the input $(N_{\text{train}}=280)$ serves as the train set and the remaining 30\% $(N_{\text{test}}=120)$ as the test set. We compare three configurations: $L=1$ with identity measurement operator $\hat{O}=\hat{I}$, $L=24$ with identity operator $\hat{I}$, and $L=24$ with a random measurement operator $\hat{A}$. Here, $L=24$ provides sufficient MC and NL, whereas $L=1$ does not.

The nonlinear transformation task is shown in Fig.~\ref{fig.task}a, where a sine wave input is mapped to a sawtooth wave target. The case of $L=24$ with $\hat{O}=\hat{I}$ achieves the lowest MSE of $6.306 \times 10^{-4}$ and $3.612 \times 10^{-4}$ on the train and test sets respectively, closely tracking the target signal. The case of $L=1$ performs significantly worse, with MSE of order $10^{-2}$, as it lacks the nonlinearity necessary to reconstruct the target and therefore returns a sine-like curve.
In Fig.~\ref{fig.task}b, the Mackey-Glass 20-step-ahead prediction task is performed. The $L=24$ configurations again outperform $L=1$ by approximately one order of magnitude in MSE. The prediction curve of $L=1$ exhibits a temporal delay relative to the target line, reflecting the absence of past input dependency. In contrast, the $L=24$ cases accurately capture the chaotic dynamics of the Mackey-Glass series.
Notably, in both tasks, the $L=24$ with random observable $\hat{A}$ achieves performance comparable to that with $\hat{O}=\hat{I}$, demonstrating that the reservoir performance is robust to the choice of measurement operator. These results verify that the extrinsically induced memory substitutes for intrinsic memory, enabling the memoryless quantum dot to successfully execute tasks demanding high memory capacity.

\begin{figure}
    \centering
    \includegraphics[width=1.0\linewidth]{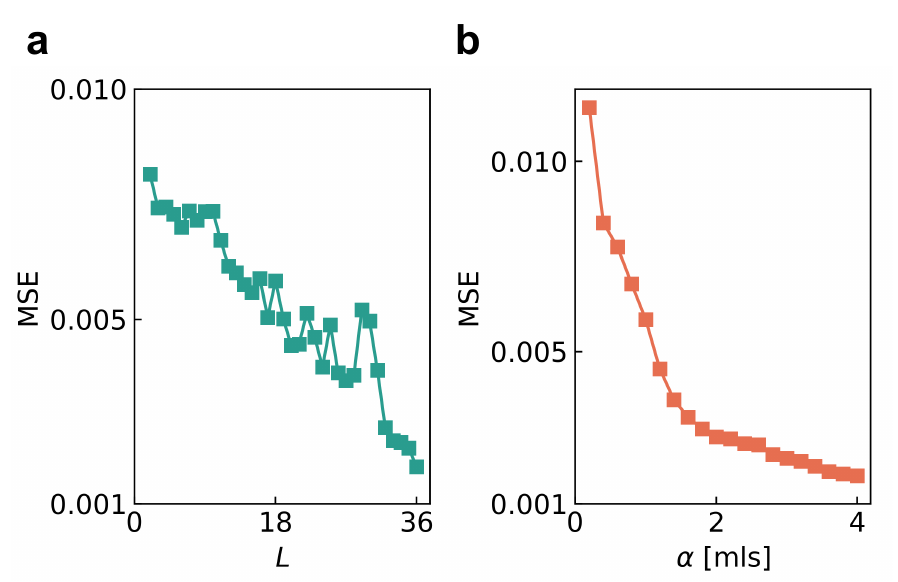}
    \caption{\textbf{MSE of the Mackey-Glass future prediction task as a function of control parameters.} (a) MSE on the test set as a function of $L$ with fixed $\alpha=1.0$. The MSE decreases monotonically with $L$, reflecting the critical role of MC. (b) MSE as a function of $\alpha$ with fixed $L=18$. Here, the improvement saturates at large values of $\alpha$, indicating the upper-bounded contribution of NL to this prediction task. Both panels use $N=400$ and $\Gamma=0.5$.}
    \label{fig.task-params}
\end{figure}
To further examine how the control parameters govern task performance, we plot the test-set MSE for the previously evaluated Mackey-Glass prediction task as a function of $L$ and $\alpha$ separately in Fig.~\ref{fig.task-params}. The MSE as a function of $L$ is plotted with fixed $\alpha=1.0$, as shown in Fig.~\ref{fig.task-params}a. The MSE systematically decreases as $L$ increases, since a larger $L$ directly extends the input history accessible to the reservoir, providing more past information to predict future states of the chaotic series.
Turning to the role of nonlinearity, the MSE as a function of $\alpha$ is shown with fixed $L=18$ in Fig.~\ref{fig.task-params}b. Here, the error also decreases with increasing $\alpha$, a parameter that primarily drives NL, indicating that the prediction task requires not only memory but nonlinearity as well. Such a requirement is consistent with the fundamental goal of the task, which is to identify the underlying nonlinear function that governs the chaotic dynamics of the Mackey-Glass series.

One distinct behavior of the $\alpha$ dependence is that the improvement saturates at large $\alpha$ and this can be understood from the eigenmode mixing picture. As $\alpha$ increases, the reservoir gains richer nonlinear capability, which is beneficial for capturing the nonlinear dynamics of the Mackey-Glass series. However, since the number of available eigenmodes is bounded by the system size, the nonlinear enrichment of the readout asymptotically reaches its maximum, and further increasing $\alpha$ yields no significant improvement in performance. In terms of reservoir metrics, the reduction in MSE with respect to $L$ and $\alpha$ reflects the respective roles of MC and NL. The saturation at large $\alpha$ suggests that the benefit of NL is upper-bounded, whereas MC continuously enhances performance for larger $L$.

\section{Homology Analysis in Hilbert Space}
\label{sec.homology}

\begin{figure*}
    \centering
    \includegraphics[width=0.8\linewidth]{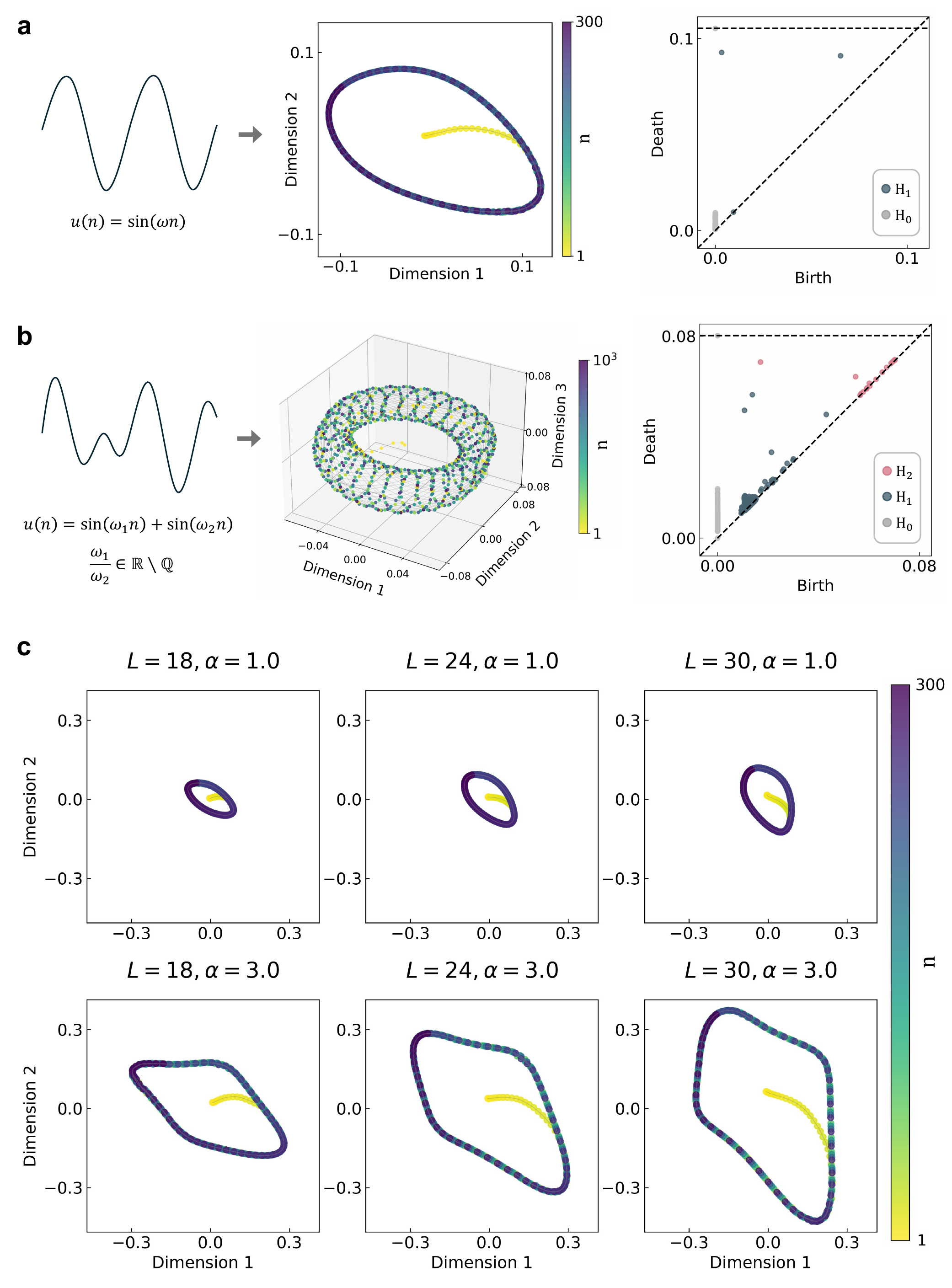}
    \caption{\textbf{Geometric analysis of reservoir state trajectories via persistent homology.} (a) When a single-frequency sine input is injected, the reservoir states trace out a closed loop in the 2D visualization space obtained via Multidimensional Scaling (MDS), colored according to time step $n$. The corresponding persistence diagram confirms this topology, showing two dominant $H_1$ points, where the one with a larger lifetime corresponds to the global loop structure while the other reflects a local feature from the early states. (b) When the input consists of two superimposed incommensurate sine waves, the trajectory expands into a torus in the 3D MDS visualization space. The persistence diagram exhibits two distinct $H_1$ points (minor and major circles) and a single $H_2$ point (cavity enclosed by the torus surface), confirming the torus topology. (c) MDS-projected trajectories for a single-frequency sine input across varying $L$ and $\alpha$. Increasing $L$ enlarges the loop while preserving its shape, whereas increasing $\alpha$ expands the loop with accompanying geometric distortion.}
    \label{fig.homology}
\end{figure*}

We focus on the result that a hysteresis loop is formed in the input-response space, as evidence of extrinsic memory. Since the reservoir's state trajectory underlies this loop structure, we analyze the geometric properties of the trajectory driven by the discrete input.
For a given readout energy $E$, we obtain a sequence of $N$ density matrices $\rho(n)$ (where $n=1, \dots, N$) expressed as:
\begin{equation}
    \rho(n) = \frac{1}{Z(n)} \sum_k \frac{\Gamma}{(E-E_k(n))^2 + \Gamma^2} \ket{\psi_k(n)}\bra{\psi_k(n)},
\end{equation}
where $Z(n)$ is a normalization factor. For the analysis of the trajectory along these mixed states, we calculate the Bures distance between all pairs of density matrices in the sequence, defined as:
\begin{equation}
    d_{n, m} = \arccos\left(\sqrt{F(\rho(n), \rho(m))}\right),
\label{eq:distance_matrix}
\end{equation}
where the fidelity is given by $F(\rho_1, \rho_2) = \left[ \text{tr}\left(\sqrt{\sqrt{\rho_1} \rho_2 \sqrt{\rho_1}}\right) \right]^2$, and $n, m \in [1, N]$.
From these pairwise calculations, an $N \times N$ distance matrix is constructed. Visualization of this trajectory is achieved through Multidimensional Scaling (MDS) \cite{torgerson1952multidimensional}, projecting the high-dimensional quantum states into a lower-dimensional visualization space. Specifically, for a 2D projection, MDS determines vectors $\mathbf{x}_1, \dots, \mathbf{x}_N \in \mathbb{R}^2$ such that the Euclidean distances approximate the quantum distances, i.e., $||\mathbf{x}_n - \mathbf{x}_m|| \approx d_{n,m}$. Note that this set of vectors is invariant to translations, rotations, and reflections. Furthermore, these projected coordinates serve strictly for visualization purpose and are not utilized in any subsequent quantitative calculations.

To quantitatively distinguish the properties of these trajectories, we employ persistent homology \cite{edelsbrunner2002topological, zomorodian2004computing} to extract topological quantities directly from the distance matrix (see Appendix~\ref{appendix.intro-homology} for more details). Persistent homology tracks the birth and death of topological features, specifically generalized holes, across varying length scales. These features are categorized by homology groups $H_k$, where $H_0$ represents connected components, $H_1$ corresponds to loops, and $H_2$ denotes cavities. We represent these features using a persistence diagram, as shown in the rightmost column of Fig.~\ref{fig.homology}a, b. In these diagrams, each topological feature is plotted as a point $(b, d)$, where $b$ and $d$ denote the diameter values at which a feature appears (birth) and disappears (death), respectively. The $0$-dimensional homology $H_0$ measures local density and clustering of the $N$ state points, which is less relevant to the emergence of the extrinsic memory loop. Our primary focus is on higher-dimensional homologies, such as $H_1$ or $H_2$, which measure the cyclic structures and loop topology of the trajectory. The stability of each homology point is estimated by its lifetime, defined as $\mathcal{L} = d-b$. Points located far from the diagonal ($d=b$) possess large lifetimes, indicating stable topological features. In contrast, points lying close to the diagonal represent topological noise, corresponding to local structures that form and disappear rapidly.

For a single-frequency sine input, the state trajectory of the reservoir projected into a 2D visualization space via MDS, alongside its persistence diagram, is illustrated in Fig.~\ref{fig.homology}a. The periodic nature of the input drives the reservoir states to trace out a closed loop, and the persistence diagram confirms this topology. There are two dominant $H_1$ points. The one with a larger lifetime corresponds to the global loop structure, while the one with a shorter lifetime reflects a local feature arising from the early states, indicated by the yellow points that have not yet settled onto the loop.
Extending this analysis to a more complex signal, the results for an input consisting of two superimposed incommensurate sine waves are presented in Fig.~\ref{fig.homology}b. From the 3D trajectory, we confirm that the reservoir expands this input into a torus structure, indicating that the system captures higher-dimensional features embedded in the input. The persistence diagram supports this result, exhibiting two distinct $H_1$ points and a single $H_2$ point, which are the topological signature of a torus. Based on these geometric analyses, we conclude that the input's geometry is successfully reconstructed within the quantum Hilbert space~\cite{ohkubo2024reservoir}.

Investigating how the control parameters shape these topological features, we display the evolution of the loop geometry for the single-frequency sine input as a function of $L$ and $\alpha$ in Fig.~\ref{fig.homology}c. At $\alpha=1.0$ (top row), increasing $L$ enlarges the loop while preserving its near-circular shape. In contrast, at $\alpha=3.0$ (bottom row), the loop expands but its shape becomes distorted, deviating from a smooth ellipse into an irregular ring. This distinction highlights that $L$ and $\alpha$ are geometrically distinct control parameters, where $L$ produces a shape-preserving enlargement of the loop while $\alpha$ introduces geometric distortion through stronger nonlinear mixing of eigenmodes.

\begin{figure}
    \centering
    \includegraphics[width=1.0\linewidth]{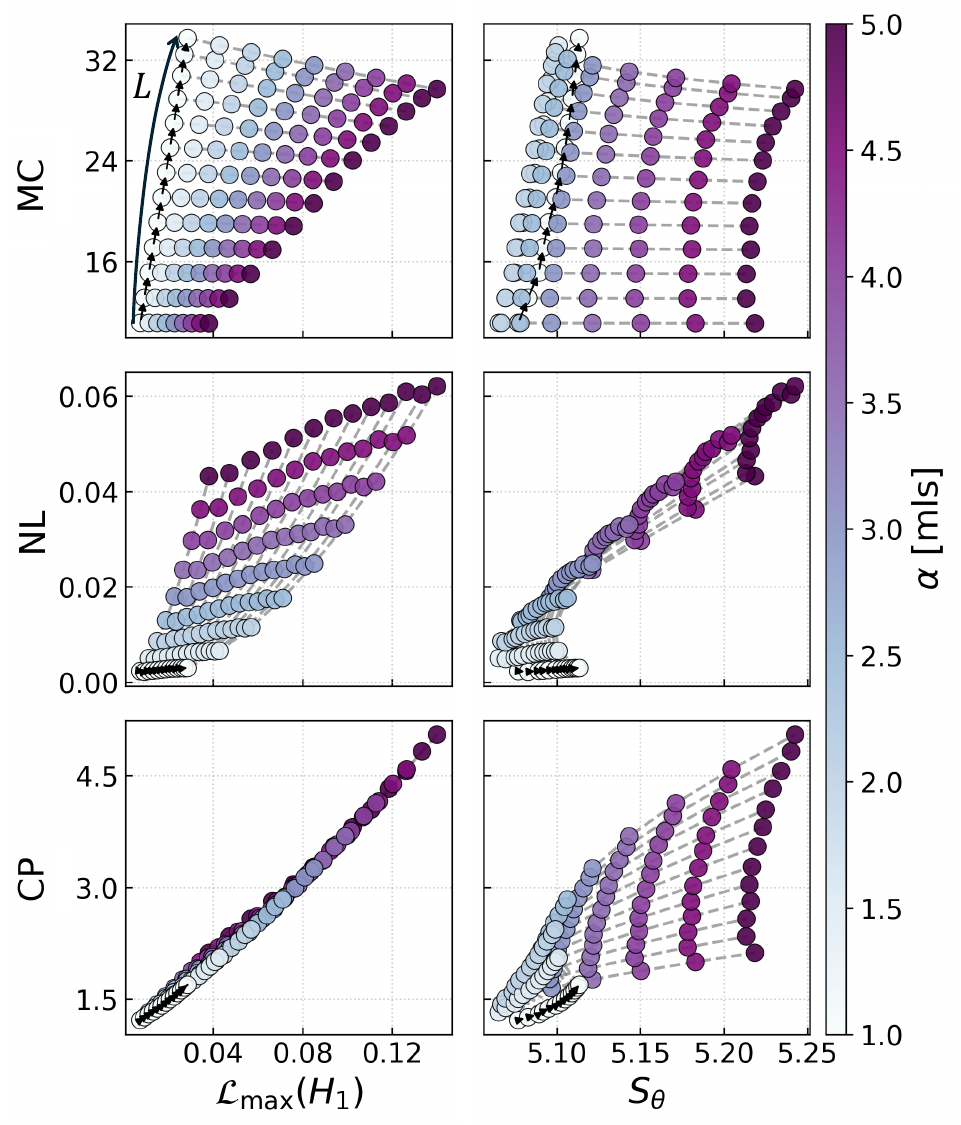}
    \caption{\textbf{Correlations between reservoir metrics and geometric metrics of the state trajectory.} Scatter plots of MC (top), NL (middle), and CP (bottom) against $\mathcal{L}_{\text{max}}(H_1)$ (left column) and $S_\theta$ (right column), computed from the distance matrix driven by a single-frequency sine input at $\Gamma=4.0$, averaged over $20$ random realizations of $\hat{H}_{\text{sys}}$. Each point corresponds to a specific $(L, \alpha)$ with $L$ ranging from 4 to 36 in steps of 2 and $\alpha$ from 1.0 to 5.0 in steps of 0.5, colored by $\alpha$. Black arrows connect points of increasing $L$ at $\alpha=1.0$, and gray dashed lines connect points of increasing $\alpha$ at fixed $L$. $\mathcal{L}_{\text{max}}(H_1)$, representing the loop size, correlates strongly with CP and NL, while the angular path length $S_\theta$ correlates with NL and remains insensitive to $L$.}
    \label{fig.homology-correlation}
\end{figure}

Now we look at the correlations between the reservoir metrics and geometric properties of the state trajectory. Since the loop scales its size or is distorted across the parameter space (see Fig.~\ref{fig.homology}c), we introduce two corresponding metrics: $\mathcal{L}_{\text{max}}(H_1)$ indicating the maximum lifetime of $H_1$, and the angular path length $S_\theta$. Here, we define $S_\theta$ as the sum of the angle $\theta_n$ at each time step (see Appendix~\ref{appendix.angular} for more details). Because we operate within a metric space projected from the Hilbert space, which does not have directional information, we calculate the magnitude of the angles of the trajectory relative to the centroid of the $N$ state points. For each $n$, the angle $\theta_n$ is defined by the triangle formed by the centroid, $n$-th point, and $(n+1)$-th point, using the distances between these three points. From this, the angular path length is calculated as $S_\theta=\frac{1}{2\pi}\sum_{n=1}^N \theta_n$.

Figure~\ref{fig.homology-correlation} shows the correlations between the two geometric features ($\mathcal{L}_{\text{max}}(H_1)$ and $S_\theta$) and the three reservoir properties (MC, NL, and CP) across the control parameter space. The single-frequency sine input is used to compute these geometric quantities, with $L$ ranging from 4 to 36 in steps of 2, and $\alpha$ from 1.0 to 5.0 in steps of 0.5. Each point corresponds to a specific $(L, \alpha)$ configuration, colored by $\alpha$. The black arrows connect points of increasing $L$ at the lowest value of $\alpha=1.0$, while the gray dashed lines connect points of increasing $\alpha$ at fixed $L$. These results are shown for $\Gamma=4.0$, where the suppression of NL and CP is enough to clearly separate their dependence on the two geometric axes, making the correlations more interpretable than at smaller $\Gamma$.

The two features show distinct correlation patterns with the reservoir metrics. $\mathcal{L}_{\text{max}}(H_1)$ exhibits a strong correlation with MC along the $L$ axis, NL along the $\alpha$ axis, and CP across both. In the top-left panel, for a fixed $L$ (gray dashed lines), increasing $\alpha$ expands $\mathcal{L}_{\text{max}}(H_1)$ while MC remains approximately constant. Conversely, following the direction of the black arrow, increasing $L$ at a fixed $\alpha$ enhances both $\mathcal{L}_{\text{max}}(H_1)$ and MC. For the NL correlation, $\alpha$ mainly drives NL upward and simultaneously enlarges the loop size, consistent with the results in Fig.~\ref{fig.homology}c. The near-perfect correlation with CP is physically intuitive, as a large loop implies that the reservoir states span a wider range of distinct configurations, directly increasing the effective dimensionality of the readout matrix, and thus CP.
The second geometric feature, $S_\theta$, reveals a correlation with NL across both axes and with CP along the $\alpha$ axis, while showing no dependence on MC. Given that the range of variation in $S_\theta$ caused by $\alpha$ is significantly greater than that caused by $L$, we confirm that $S_\theta$ effectively captures the geometric distortion of the trajectory. In the top-right panel, MC scales almost vertically along the $L$ axis, which indicates that introducing extrinsic memory via $L$ does not impose additional distortion on the trajectory. Consequently, the loop structure becomes topologically more stable as MC increases, because a higher MC yields a larger loop size while maintaining a constant level of distortion. Furthermore, the correlation between $S_\theta$ and NL reflects the fact that NL arises not merely from the size of the loop but from the irregularity of the trajectory, as a distorted loop introduces angular variations unrelated to the input signal.

\section{Discussion}
\label{sec.discuss}

We set out to ask whether a memoryless system can serve as an efficient reservoir when its memory is drawn entirely from outside, and our results show that it can. Our extrinsic memory method rests on the space-time tradeoff, spending spatial degrees of freedom to buy back the retention time that a memoryless system cannot provide. Since the tradeoff is a computational principle, not a material property, the scheme could even carry over to classical systems. The Mackey-Glass future prediction task then confirms that this memory does work equivalently to intrinsic memory during computation, and such validation provides evidence that the proposed memory scheme widens the class of usable reservoirs.

We choose a quantum dot as our reservoir where its discretized energy levels supply a rich nonlinear response, but this property need not be intrinsic. Nonlinearity can equally be introduced from outside the reservoir, as in the polynomial feature expansion~\cite{Gauthier2021}. When both memory and nonlinearity are brought externally, the computation is totally implemented via software, and the feature nodes grow in number polynomially with the delay length $L$ which then enter the linear regression to be trained. Because we instead let the physical system perform the nonlinear mixing, our readout dimension is set by the number of measurement regions, which is proportional to the system size, and the training cost grows only linearly with $L$. A reservoir with intrinsic nonlinearity therefore keeps the computation cheap as the required memory depth increases.

The geometric analysis we introduce offers a way to read the computational properties of a reservoir directly from its state trajectory. When an input signal is applied via the proposed injection scheme, it drives the quantum system to form a hysteresis loop in Hilbert space, with the three reservoir metrics leaving distinct fingerprints. Memory capacity opens the loop and enlarges it without deforming it, nonlinearity distorts its shape, and complexity follows the range it spans. The geometry reports on the input as well. We show that the reservoir unfolds a one-dimensional signal built from two incommensurate sine waves into a torus whose cavity appears as a two-dimensional hole $H_2$ in Hilbert space, so the hole dimension reflects the effective dimensionality of the input. The geometry therefore speaks to both reservoir properties and the input's dimensionality. For the reservoir side, this means its capabilities are read from the states alone---with no weights fitted---which suggests a criterion to assess candidate reservoirs before committing them to a benchmark task.

Several features of our scheme ease its experimental realization. The readout in Eq.~\ref{eq:readout} is an expectation value, which in an isolated system would have to be built up from repeated runs. In practice, perfectly isolating a quantum reservoir is challenging, so the system is inevitably coupled to its environment. However, the resulting decoherence performs this averaging on its own, and therefore the measured signal converges to the energy-resolved values $r_m(n; E)$. When the observable is the identity, $\hat{O}=\hat{I}$, these values represent a spatially resolved local density of states (LDOS). This is closely analogous to a scanning tunneling microscopy measurement in which the tunneling current at each tip position is proportional to the LDOS at the bias energy $E$, with $\Gamma$ playing the role of the tip-sample broadening. The measurement itself need not be finely engineered, since a randomly generated observable $\hat{A}$ performs comparably to the identity on both tasks. Finally, the computation runs entirely in discrete time, so that the input rate is set by the experimenter rather than by the device's relaxation time, and the computation can be halted and resumed at any step.



\begin{acknowledgements}
We thank Hidekazu Kurebayashi, Rosa Lopez, and Junhyeon Bae for useful discussions and careful reading of the manuscript.
B.K. and K.W.K. acknowledge financial support from the Basic Science Research Program through the National Research Foundation of Korea (NRF) funded by the Ministry of Education (No. RS-2025-00521598). B.K. was supported by the Chung-Ang University Graduate Research Scholarship in 2026. S.J. and K.W.K. were supported by the National Research Foundation of Korea (NRF) (Grant No. RS-2020-NR049536).
\end{acknowledgements}

\bibliography{reference.bib}

\begin{figure*}[t!]
    \centering
    \includegraphics[width=1.0\linewidth]{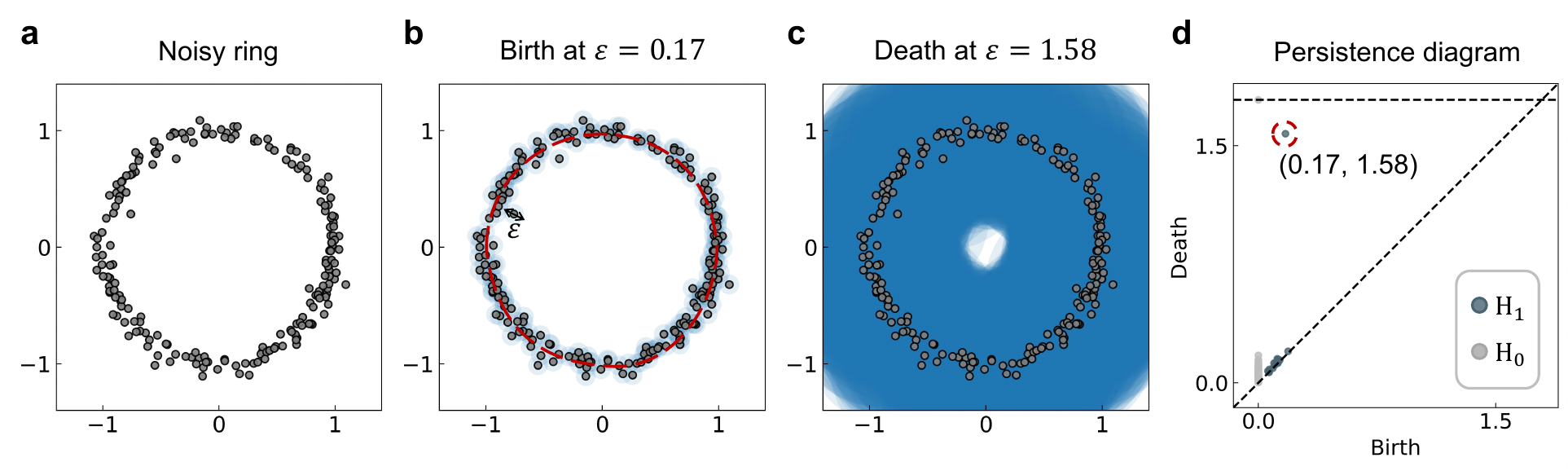}
    \caption{\textbf{A simple illustration of persistent homology on a noisy ring}. (a) 200 points are randomly sampled on a circle of radius $r=1$ with small radial noise. (b) As the diameter $\varepsilon$ of the disks (blue) centered at each point grows, the dominant loop (red dashed line) closes and is born at $\varepsilon=0.17$. (c) The loop dies at $\varepsilon=1.58$, where the disks have grown large enough to connect points across the ring. (d) The persistence diagram shows a single long-lived $H_1$ point at $(0.17, 1.58)$ that captures the major loop, while the remaining $H_1$ points lie near the diagonal and represent local noise in the data.}
    \label{fig.intro-homology}
\end{figure*}

\newpage
\appendix
\newpage

\section{Introduction to Persistent Homology}
\label{appendix.intro-homology}

In this work, our main goal is to check whether the extrinsic memory forms a hole in the system's quantum Hilbert space, which would be the evidence of hysteresis-like behavior. We employ homology, a generalized hole, to measure geometrical structure of the quantum trajectory during the computation. 
To extract the homology information from the trajectory, we use the persistent homology method. The basic building block is a simplex, the generalization of a point (0-simplex), an edge (1-simplex), and a triangle (2-simplex) to arbitrary dimension. Any geometric shape can be approximated by gluing such simplexes together along shared faces, and persistent homology constructs this approximation directly from the data, expanding it as a scale parameter $\varepsilon$ grows. Two points within $\varepsilon$ are joined by an edge, three points within $\varepsilon$ are filled by a triangle, and so on. Equivalently, one can picture a disk of diameter $\varepsilon$ centered at each point (blue disks in Fig.~\ref{fig.intro-homology}b), so that an edge forms precisely when two disks overlap. As $\varepsilon$ increases, topological features appear and disappear, and the persistence diagram records this information. More persistent features lie far from the diagonal line (birth$=$death) and represent global geometric properties, whereas less persistent features lie close to the diagonal and represent local properties such as artifacts or noise.

For example, consider the noisy ring data, which has a loop structure $(H_1)$, shown in Fig.~\ref{fig.intro-homology}a. Fig.~\ref{fig.intro-homology}b-c show the birth and death of the most dominant loop structure. The loop is born at $\varepsilon=0.17$, where the edges first connect the points all the way around the ring into a closed loop. It dies at $\varepsilon=1.58$, where triangles fill in the region the loop encloses. Through the persistence diagram in Fig.~\ref{fig.intro-homology}d, we can read all the homology in the data. Because the ring data lie in the two-dimensional plane, only $H_0$ and $H_1$ can be nontrivial. For $H_0$, every point appears at $\varepsilon \rightarrow 0^+$, and each $H_0$ class dies when its component merges into another. The single component that eventually contains the entire point cloud never merges and therefore never dies. This class is drawn on the horizontal dashed line, which marks death$=\infty$. For $H_1$, the points near the diagonal (birth $\simeq$ death) indicate local noise, while the single point far from it (birth $\ll$ death) reveals the dominant 1d-loop structure.

\section{Estimation of the Angular Path Length}
\label{appendix.angular}
Here we provide detailed information on the angular path length $S_\theta$ that we introduce to quantify how much the quantum trajectory rotates during the computation. To estimate this rotation we first need a reference point, for which we take the centroid of the reservoir states. Using the distance matrix defined in Eq.~\ref{eq:distance_matrix}, the centroid point is defined as the distance $d_{c, i}$ between the centroid and the $i$-th density matrix of the reservoir as \cite{Havel1983}:
\begin{equation}
    d_{c,i}^2 = \frac{1}{N}\sum_{j=1}^N d_{i,j}^2 - \frac{1}{N^2}\sum_{k>j}^N d_{j,k}^2,
\end{equation}
where $N$ is the number of reservoir states (equal to the input length).
At each time step $n$, the angle $\theta_n$ is formed by the triangle connecting the centroid and the consecutive states $n$ and $n+1$. Using the law of cosines, this angle is given by:
\begin{equation}
    \theta_n = \arccos(\frac{d_{c,n}^2 + d_{c, n+1}^2 - d_{n, n+1}^2}{2d_{c,n}  \ d_{c, n+1}}).
\end{equation}
Note that because the distances are defined within a metric space without any specified orientation, $\theta_n$ is an undirected, non-negative angle.
The angular path length $S_\theta$ is then the cumulative sum of these angles, as defined in the main text.

\section{Comparative Analysis of Spatial-Temporal Input Mappings}
\label{appendix.mappings}
\begin{figure*}
    \centering
    \includegraphics[width=1.0\linewidth]{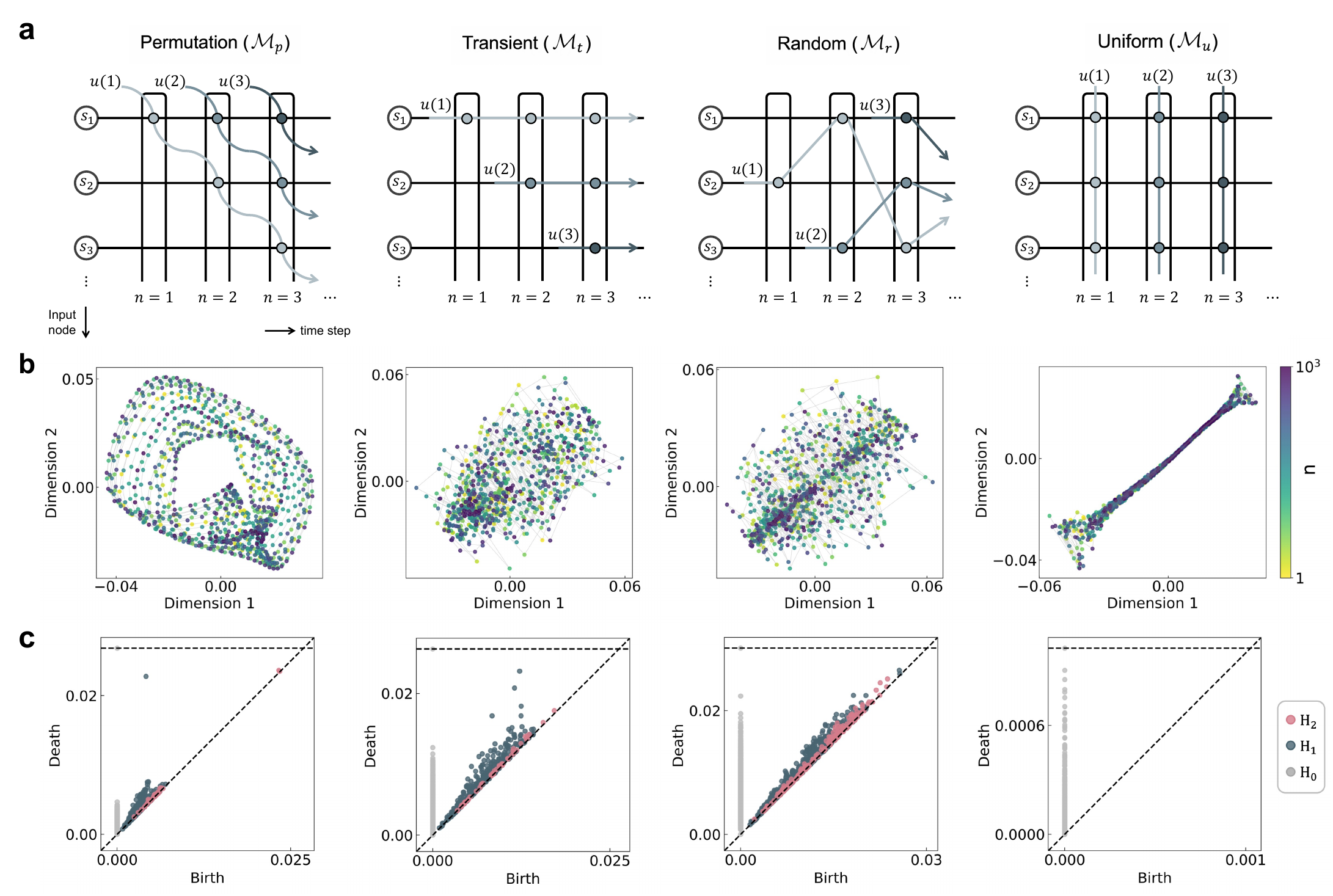}
    \caption{\textbf{Spatial-temporal input mappings and their topological features on quantum trajectories.} (a) The signals $u(1)$, $u(2)$, and $u(3)$ represent the inputs at time steps $n=1$, $2$, and $3$, respectively, and $s_j$ denotes the mapped input fed into node site $j$ via the injection map $\mathcal{M}$. In permutation mapping ($\mathcal{M}_p$), the signal shifts sequentially across nodes. In transient scheme ($\mathcal{M}_t$), once a specific input is assigned to a certain node, it remains there for $L$ steps and then is overwritten by the new input. In random scheme ($\mathcal{M}_r$), the destination node is chosen randomly at each time step. In these three schemes, each signal is retained for a duration of $L$. For comparison, we lastly introduce a memoryless mapping $\mathcal{M}_u$ that maps a non-delayed input $u(n)$ to all nodes, retaining no past information.
    (b) Quantum state trajectories are projected into a 2D space via the MDS method. The reservoir consists of $N_s=36$ sites with $L=9$ input nodes, driven by $\text{MG}(n)$ as an input signal of length $N=1,000$. $\mathcal{M}_p$ forms distinct, closed loops indicating a successful reconstruction of the signal's geometry. In contrast, $\mathcal{M}_t$ exhibits an unstable central hole, and $\mathcal{M}_r$ shows no consistent structure. The memoryless mapping $\mathcal{M}_u$ collapses the trajectory onto a diagonal path. 
    (c) Persistence diagrams reveal the topological stability of the state trajectories. The diagrams plot the birth versus death of homology features, where the vertical distance from the diagonal ($\mathcal{L}=\text{death}-\text{birth}$) represents the feature's stability. The analysis focuses on 1-dimensional homology $H_1$ (loops). $\mathcal{M}_p$ exhibits a long-lived $H_1$ point, confirming the formation of a stable global loop. The hierarchy of the $H_1$ maximum lifetimes, $\mathcal{L}_{\max}(\mathcal{M}_p) > \mathcal{L}_{\max}(\mathcal{M}_t) > \mathcal{L}_{\max}(\mathcal{M}_r)$, demonstrates that the input injection scheme is critical for stable topological encoding. $\mathcal{M}_u$ lacks loop features, indicating a topologically trivial structure.}
    \label{fig.mappings}
\end{figure*}
Our main idea is to introduce memory externally by injecting $L$ multi-delayed input signals into $L$ spatial input nodes on the lattice. In the main text, the injection protocol of Eq.~\ref{eq:permute} maps $(i-1)$-step delayed input to the $i$th input node, where $i\in[1, L]$. 
Mathematically, we can generalize the injection map $\mathcal{M} \in \mathbb{R}^{L\times L}$ as $\mathbf{s}_n = \mathcal{M} \mathbf{u}_n$, where $\mathbf{u}_n = [u(n), \dots, u(n-L+1)]^T \in \mathbb{R}^L$ and $\mathbf{s}_n \in \mathbb{R}^L$ is the node-input vector (component $j$ feeds node $j$). Here we introduce four distinct mappings which differ in how the multi-delayed signals are distributed across the input nodes: permutation $(\mathcal{M}_p)$, transient $(\mathcal{M}_t)$, random $(\mathcal{M}_r)$, and uniform $(\mathcal{M}_u)$, as shown in Fig.~\ref{fig.mappings}a. Note that $\mathcal{M}_p$ corresponds to the mapping used in the main text. In the following we demonstrate that $\mathcal{M}_p$ yields the most persistent one-dimensional loop, signaling the largest extrinsic memory, as revealed through a quantum trajectory approach.

At each time step, the input Hamiltonian $\hat{H}_{\text{in}}(n)$ is defined as:
\begin{equation}
\hat{H}_{\text{in}}(n) = \alpha \sum_{j=1}^L s_j(n) c_j^\dagger c_j,
\end{equation}
where $s_j(n) = (\mathcal{M}\mathbf{u}_n)_j$ is the input fed to the node site $j$, as set by the injection map $\mathcal{M}$. Here, the inputs are encoded into an extra onsite potential.
For $\mathcal{M}_p$, each input is first injected at the 1st node and then advances one node per step until it reaches the $L$th node as its last stop, so that over its lifetime each input is permuted across the whole input-node space. This corresponds to $\mathcal{M}_p = I$, the identity matrix, giving $s_j(n)=u(n-j+1)$ for $j=1, \dots, L$.
For $\mathcal{M}_t$, by contrast, each node holds its value for $L$ steps, with the refresh phases staggered so that exactly one node updates per step. The value resident at any node is therefore transient, persisting only for $L$ steps before being overwritten. This is described mathematically by $s_j(n)=u(n-d_j(n))$, where the age of the held sample is $d_j(n)=(n+j-1) \bmod{L}$.
For $\mathcal{M}_r$, the components of $\mathbf{u}_n$ are assigned to nodes through a permutation drawn at random, independently at every time step. That is, the matrix $\mathcal{M}_r$ keeps changing randomly in time while preserving the full input history.
Finally, for $\mathcal{M}_u$ we set $(\mathcal{M}_u)_{jk} = \delta_{k, 1}$, so that only the current input $u(n)$ is uniformly assigned to all input nodes. 
Note that $\mathcal{M}_p, \mathcal{M}_t$, and $\mathcal{M}_r$ all utilize the identical input history, whereas $\mathcal{M}_u$ uses the current input only, which indicates that it carries no extrinsic memory. Even among the three memory-inducing mappings, however, the amount of memory they supply differs, as we analyze in Appendix~\ref{appendix.synchronize}.

In Fig.~\ref{fig.mappings}b, we visualize the quantum trajectories for each of the four mappings when the Mackey-Glass (MG) chaotic time series is injected.
In the case of $\mathcal{M}_p$, the state traverses distinct loops. This topology indicates that the mapping captures the quasi-periodicity of $\text{MG}(n)$, leading to the formation of multiple loops. Furthermore, the state follows distinguishable paths within these loops, reflecting the system's ability to resolve the chaotic property of the input through the quantum dot's inherent nonlinearity.
For $\mathcal{M}_t$, while the state moves along a distinct path over time, it fails to form a coherent trajectory structure comparable to that of $\mathcal{M}_p$. Moreover, the $\mathcal{M}_t$ trajectory exhibits an ill-defined hole at its center, whereas $\mathcal{M}_p$ maintains a clear, stable one.
The trajectory of $\mathcal{M}_r$ exhibits neither a clear central hole nor a consistent path. Instead, a significant number of points are observed to cluster along the diagonal. 
Finally, for $\mathcal{M}_u$, the system retraces almost the same diagonal path for identical input values, a behavior indicative of a memoryless response.
Based on these geometric analyses, the reservoir's response dynamics can be interpreted as a mapping from the temporal input space to the geometry within the quantum Hilbert space. By employing schemes such as $\mathcal{M}_p$, we effectively reconstruct the geometry of the signal's history \cite{ohkubo2024reservoir} within the Hilbert space. In contrast, $\mathcal{M}_u$ fails to achieve this, resulting in a collapse of the geometry. 

\begin{figure*}
    \centering
    \includegraphics[width=0.7\linewidth]{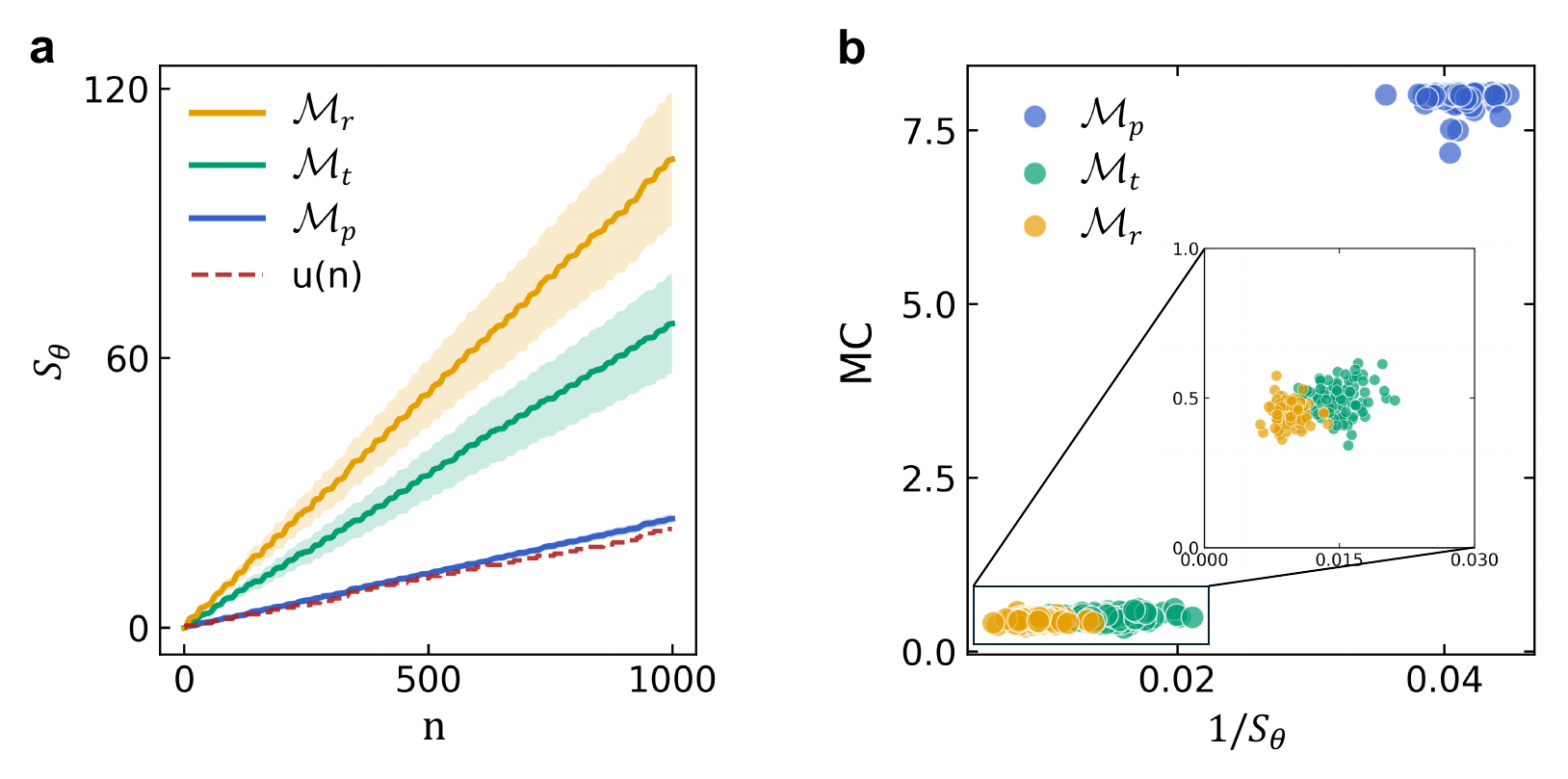}
    \caption{\textbf{Correlation between angular path length and $\text{MC}$ across three memory-inducing mappings.} (a) Angular path length $S_\theta$ as a function of time step $n$ for the three mappings $\mathcal{M}_p, \mathcal{M}_t, \mathcal{M}_r$, together with the input signal $u(n)=\text{MG}(n)$ as a reference. Shaded regions represent the standard deviation across $100$ random realizations of $\hat{H}_{\text{sys}}$. $S_\theta$ reflects the system's ability to capture the periodicity of $u(n)$. A slope closer to that of the input signal indicates better geometric reconstruction of the input in the quantum trajectory. $\mathcal{M}_p$ tracks the input well, whereas $\mathcal{M}_t$ and $\mathcal{M}_r$ exhibit a steeper slope than $u(n)$, indicating that they fluctuate in Hilbert space more rapidly than the input's inherent oscillation. 
    (b) Total MC for $\hat{O}=\hat{I}$ versus the inverse path length $1/S_\theta$. Each mapping is
represented by $100$ points, each corresponding to a different random realization of $\hat{H}_{\text{sys}}$.}
    \label{fig.synchronize}
\end{figure*}

The persistence diagrams shown in Fig.~\ref{fig.mappings}c reveal the geometric properties of the trajectories. For every mapping, all $H_2$ points lie close to the diagonal line, which indicates that $H_2$ class is trivial here due to the input signal's dimensionality. Consequently, what is relevant to the extrinsic memory is $H_1$.
For $\mathcal{M}_p$, the diagram exhibits a single long-lived $H_1$ point. This confirms the existence of a stable loop structure in the Hilbert space, and the trajectory is effectively embedded into a two-dimensional plane.
Comparing the mappings, we observe a hierarchy in the maximum lifetime of $H_1$: $\mathcal{L}_{\max}(\mathcal{M}_p) > \mathcal{L}_{\max}(\mathcal{M}_t) > \mathcal{L}_{\max}(\mathcal{M}_r)$. In contrast, $\mathcal{M}_u$ does not generate any $H_1$ features, indicating a topologically trivial path. It is important to note that these three methods utilize the same set of information at every time step, differing only in the spatial assignment of the input nodes relative to the signal's temporal history. This suggests that the spatial arrangement is the critical factor in the formation of stable topological loops. Therefore, through $\mathcal{M}_p$, we extrinsically induce a hysteresis loop within the Hilbert space, despite the absence of hysteresis in the system's inherent dynamics.

\section{Quantum Trajectory Synchronization with Input Signal}
\label{appendix.synchronize}
While the homology analysis in Appendix~\ref{appendix.mappings} confirms the reservoir's ability to capture the input's quasi-periodicity, it offers limited insight into the dynamical details within the loop structure. Crucially, here we argue that the generation of extrinsic MC is not solely determined by the geometric size of the loop, corresponding to $\mathcal{L}_{\text{max}}$, but rather by the formation of a coherent circular trajectory that couples with the inherent oscillation of the input. We employ the angular path length $S_\theta$ to analyze the dynamical coherence. 

The temporal evolution of the path length $S_\theta$ for the three memory-inducing mappings and input $u(n) = \text{MG}(n)$ is plotted in Fig.~\ref{fig.synchronize}a. We observe that the slope of the $\mathcal{M}_p$ closely aligns with that of the input, whereas $\mathcal{M}_t$ and $\mathcal{M}_r$ exhibit steeper slopes. This indicates that when $\mathcal{M}_p$ is applied, the state trajectory evolves in synchrony with the inherent oscillation frequency of $u(n)$. In contrast, the other methods fluctuate more rapidly than the input's oscillation. This dynamic mismatch hinders the formation of a clear hysteresis loop within the Hilbert space. Moreover, the narrow shaded regions, representing the standard deviation across random realizations of the system Hamiltonian, demonstrate that $\mathcal{M}_p$ achieves this synchronization robustly, regardless of the specific system configuration.

In Fig.~\ref{fig.synchronize}b, the correlation between the inverse angular path length $1 /S_\theta$ and total $\text{MC}$ for $\hat{O}=\hat{I}$ is illustrated. The data reveal a clear separation between the mappings where $\mathcal{M}_p$ forms a distinct cluster in the high-capacity region, whereas $\mathcal{M}_t$ and $\mathcal{M}_r$ are confined to the low-capacity region as detailed in the inset. This positive correlation confirms that suppressing excessive state fluctuations, manifested as a larger $1/S_\theta$, is essential for maximizing memory capacity. In other words, as the state trajectory deviates from the input's inherent timescale and fluctuates more rapidly, the system's ability to retain temporal history significantly degrades.

\section{Scaling Behavior of Extrinsic Memory}
\label{appendix.scaling}
\begin{figure*}
    \centering
    \includegraphics[width=0.8\linewidth]{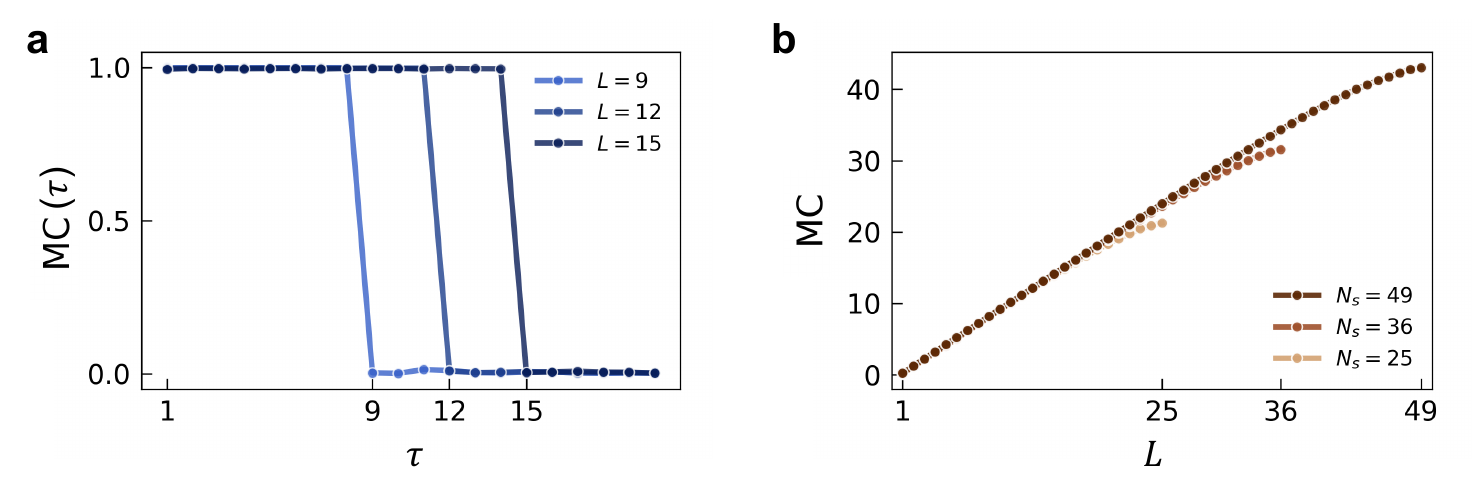}
    \caption{\textbf{Scaling behavior of the extrinsic memory with respect to $L$ and $N_s$ for $\mathcal{M}_p$.} (a) Delay-specific memory capacity $\text{MC}(\tau)$ as a function of $\tau$ across various numbers of input nodes $L$. Increasing $L$ extends the effective relaxation time, demonstrating that extrinsic MC is tunable via scaling $L$. 
    (b) Total memory capacity $\text{MC}$ versus $L$. A larger system size $N_s$ expands the maximum limit of the MC, with the realized capacity determined by $L$. All calculations were performed with $\hat{O}=\hat{I}$, $\Gamma=0.5$ and $\alpha=1.0$.}
    \label{fig.scaling}
\end{figure*}
The scaling behavior of extrinsic MC for $\mathcal{M}_p$ by varying the number of input nodes $L$ is displayed in Fig.~\ref{fig.scaling}a. We define the summand in Eq.~\ref{eq:MC} as $\text{MC}(\tau)$, which quantifies the memory capacity for a specific delay step $\tau$:
\begin{equation}
    \text{MC}(\tau) = \frac{\text{cov}^2(\hat{\mathbf{u}}_{\tau},\, \mathbf{u}_{\tau})}{\sigma^2(\hat{\mathbf{u}}_{\tau})\sigma^2(\mathbf{u}_{\tau})},
\end{equation}
where $\text{MC}(\tau)$ is bounded within $[0, 1]$, and the total memory capacity is given by $\text{MC}=\sum_{\tau=1}^k \text{MC}(\tau)$. For $L=9, 12,$ and $15$, $\text{MC}(\tau) \approx 1$ at $\tau <L$ and then abruptly drops to zero at $\tau=L$, which indicates that past inputs up to $u(n-L+1)$ are preserved. Therefore, this spatial control over memory retention allows us to systematically engineer the system to emulate or even surpass conventional reservoirs that are constrained by their intrinsic memory.
Furthermore, as shown in Fig.~\ref{fig.scaling}b, this extrinsic MC is subject to an upper-bound proportional to the system size $N_s$. Since the system is memoryless, MC is strictly zero when $L=1$, regardless of the system size. However, as $L$ increases, the total MC initially grows linearly with a unit slope (a slope of $1$) across all $N_s$. As $L$ approaches the system size, MC saturates, reaching a distinct upper limit determined by $N_s$.

\end{document}